\documentclass[conference]{IEEEtran}
\usepackage{cite} 
\usepackage{amsmath}
\usepackage{graphicx}
\usepackage{balance}

\ifCLASSINFOpdf
\else
\fi
\author{
    \IEEEauthorblockN{Somayeh Kafaie\IEEEauthorrefmark{1}\IEEEauthorrefmark{2}, Yuanzhu Chen\IEEEauthorrefmark{1}\IEEEauthorrefmark{3}, Mohamed Hossam Ahmed\IEEEauthorrefmark{2}, Octavia A. Dobre\IEEEauthorrefmark{2}}
    \\\
    \IEEEauthorblockA{\IEEEauthorrefmark{1}Wireless Networking and Mobile Computing Laboratory}
    \IEEEauthorblockA{\IEEEauthorrefmark{2}Faculty of Engineering and Applied Science, Memorial University of Newfoundland, St. John's, NL A1B 3X5, Canada}
    \IEEEauthorblockA{\IEEEauthorrefmark{3}Department of Computer Science, Memorial University of Newfoundland, St. John's, NL A1B 3X5, Canada}
}

\begin{document}
%
\title{Network Coding with Link Layer Cooperation in Wireless Mesh Networks}



%


\maketitle

\begin{abstract}
In recent years, network coding has emerged as an innovative method that helps wireless network approaches its maximum capacity, by combining multiple unicasts in one broadcast. However, the majority of research conducted in this area is yet to fully utilize the broadcasting nature of wireless networks, and still assumes fixed route between the source and destination that every packet should travel through. This assumption not only limits coding opportunities, but can also cause buffer overflow in some specific intermediate nodes. Although some studies considered scattering of the flows dynamically in the network, they still face some limitations. This paper explains pros and cons of some prominent research in network coding and proposes FlexONC (Flexible and Opportunistic Network Coding) as a solution to such issues. The performance results show that FlexONC outperforms previous methods especially in worse quality networks, by better utilizing redundant packets spread in the network.
\end{abstract}


%
\IEEEpeerreviewmaketitle

\section{Introduction}
\begin{figure}[ht]
\centering
\includegraphics[scale=0.48]{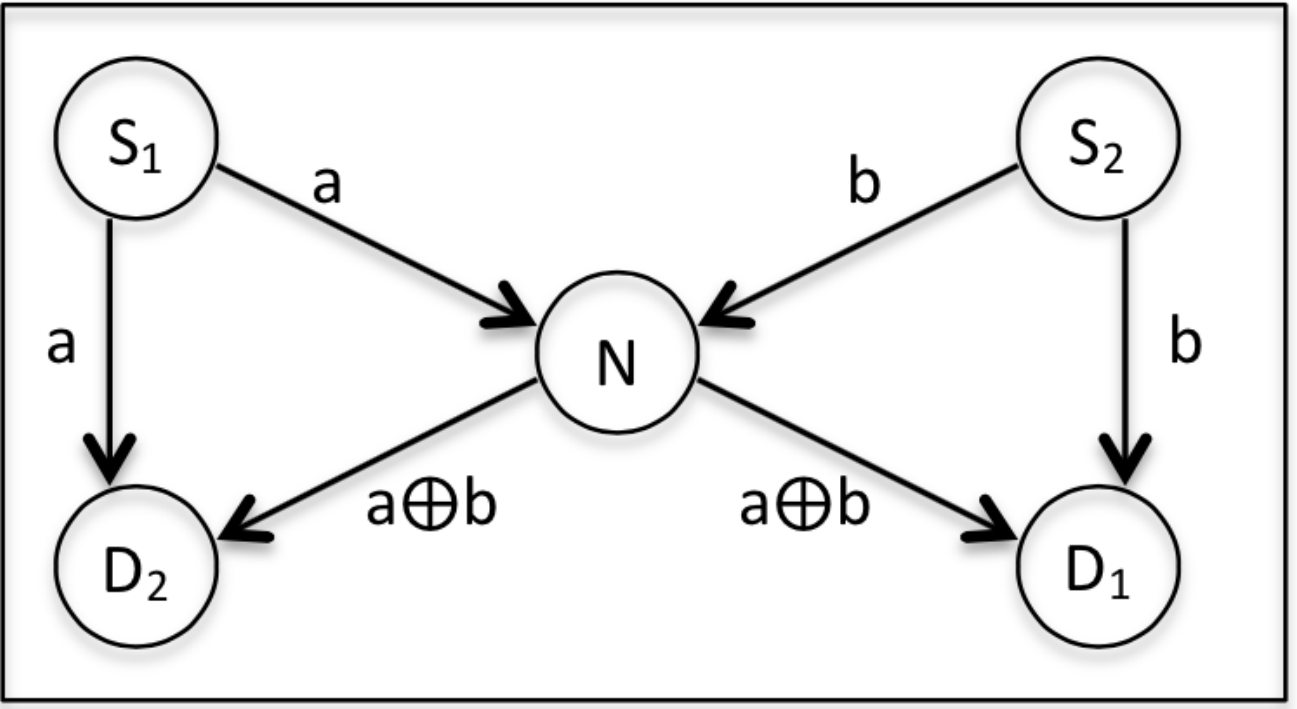}
\caption{X-topology showing how network coding improves throughput.}
\label{NCidea}
\end{figure}

Network coding represents an innovative idea introduced by Ahlswede et al. \cite{NC} in 2000 to increase the transmission capacity of the network, as well as its robustness. One of the most popular examples showing the gain behind network coding is the X-topology in Fig. \ref{NCidea}, where $S_{1}$ has packet $a$ for $D_{1}$, and $S_{2}$ has packet $b$ for $D_{2}$. While in the traditional network, four transmissions are required to deliver packets $a$ and $b$ to their final destinations, network coding decreases this number to three. Specifically, intermediate node $N$ mixes packets $a$ and $b$ together and sends $a \oplus b$. Then, $D_{1}$ ($D_{2}$), which has already overheard $b$ ($a$), is able to decode its own packet $a$ ($b$).

COPE \cite{COPE} is one of the first methods that realize this idea in practical scenarios. Whenever an intermediate node receives packets from different flows, it encodes them if it is likely that the next hops of the native packets combined in the coded packet are able to decode this packet and retrieve the original content. However, coding opportunities in COPE are restricted only to joint nodes that receive packets from multiple flows. Therefore, to provide more coding opportunities, COPE needs more packets to arrive at the same node. However this traffic concentration may overload intermediate nodes, and cause longer delay, buffer overflow, and channel contention.

As a solution to this problem, BEND \cite{BEND} applies network coding while trying to avoid traffic concentration. By taking advantage of the broadcasting nature of wireless networks, BEND allows all receivers of the packet, in addition to the intended next hop specified by the routing protocol, to help in mixing and forwarding the packet if they believe they can be helpful. However, these non-intended forwarders (i.e., the receivers of the packet which are not specified as the next hop in the route defined by the routing protocol) are allowed to mix and forward only received native packets. In fact, if they receive a coded packet, they just discard it, even if they are able to decode the received packet. This restriction not only limits the number of coding opportunities in the network but also increases the number of retransmissions.

To better utilize the broadcasting nature of wireless networks, we introduce FlexONC (Flexible and Opportunistic Network Coding) which adds more flexibility to previous methods like COPE and BEND by allowing non-intended forwarders to help in decoding in addition to encoding and forwarding. The main contributions of FlexONC are as follows: 1) More diffusion gain since more packets (i.e., coded and native packets) can be forwarded by a node other than their intended forwarder; 2) Faster packet delivery to the final destination because even if the intended forwarder does not receive the packet or cannot decode the received coded packet, some non-intended forwarders can still help; 3) More coding opportunities as non-intended forwarders are eligible to receive and probably decode coded packets and consider them as candidates to be mixed with other packets.

The rest of the paper is organized as follows. Related research on network coding, especially COPE and BEND, is discussed in Section 2. Section 3 provides an example to show the effectiveness of FlexONC and describes its objectives and challenges. Section 4 presents the implementation details of FlexONC. In Section 5, some features of FlexONC are discussed by emphasizing some pros and cons. Finally, Section 6 concludes the paper and provides ideas to extend FlexONC in future research.

\section{Background and Related Work}
In recent years, a significant amount of research has been conducted to explore the effect of network coding in different scenarios and improve the network performance by mixing packets in intermediate nodes before forwarding. Even in some publications, network coding is used to mix signals instead of bits \cite{Katti:PHYNC2007, PNC, MIXIT}.

In general, two different types of network coding can be applied, namely intra-flow and inter-flow network coding. While in the former, nodes mix packets of the same flow to increase the robustness \cite{MORE, E-NCP, ICEMAN2}, in the latter packets of different flows are mixed to reach the maximum capacity of the network \cite{COPE, BEND, Widmer+NC2005}.

COPE is one of the prominent examples of inter-flow network coding. In COPE, a node combines the packets with different next-hops when in the combined packet 1) for each next-hop there is at most one packet, and 2) the combined packet contains $P_{1}$, $P_{2}$, ..., $P_{n-1}$, the corresponding packet of $n$th next-hop will be added, if this next-hop has already received $P_{1}$, $P_{2}$, ..., $P_{n-1}$. However, the improvement of throughput in COPE depends on the traffic pattern. In fact, it limits coding opportunities because coding can be accomplished only at joint nodes.

A variety of improvements over COPE have been put forward \cite{NCSurvey2011}. BEND, as an advancement of COPE, introduces another type of gain, referred to as the diffusion gain, which is the benefit of being able to scatter flows through multiple forwarders dynamically. In BEND, each node has three queues: $Q_{1}$ for intended native packets, $Q_{2}$ for overheard native packets, and \emph{mixing-Q} for coded packets. A node can combine two packets if the next hop of the first packet is the previous hop of the second packet or one of its neighbors, and vice versa.

To avoid traffic concentration in BEND, a non-intended forwarder may receive a native packet and mix and forward it on behalf of the intended forwarder. To do so, BEND includes a \emph{second-next-hop} field in native packets, which is set by the intended forwarders. As such, when a non-intended forwarder receives a native packet, it can find the address of the next hop in the \emph{second-next-hop} field. However for coded packets, the \emph{second-next-hop} field does not present the correct address. Therefore, non-intended forwarders must drop coded packets since they do not know the address of the next hop from the intended forwarder to the destination.

\begin{figure}[ht]
\centering
\includegraphics[scale=0.48]{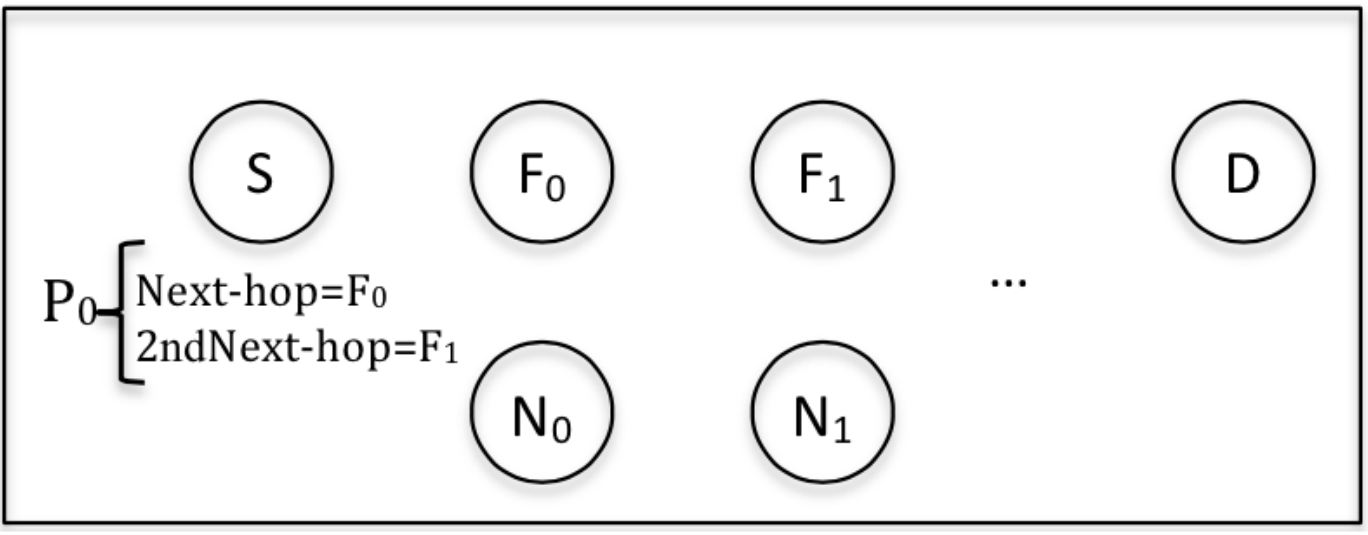}
\caption{In BEND, non-intended forwarders drop coded packets.}
\label{SNH}
\end{figure}

To illustrate the idea, let us assume in Fig. \ref{SNH} that the source $S$ sends a packet $P_{0}$ to $D$. Based on the information provided by the routing protocol, it fills the \emph{next-hop} and \emph{second-next- hop} fields with $F_{0}$ and $F_{1}$, respectively. We assume that $F_{0}$ fails to receive the packet, and $N_{0}$ overhears it. In addition, $N_{0}$ can mix $P_{0}$ with packet $P_{1}$ in its buffer. Based on $P_{0}$'s header, $N_{0}$ sets the new \emph{next-hop} field with the current \emph{second-next- hop} field, $F_{1}$. However, since it transmits a coded packet (i.e., $P=P_{0} \oplus P_{1}$), it does not change the \emph{second-next-hop} field in $P_{0}$. Now, if $F_{1}$ receives and decodes $P_{0}$ successfully, it can run the routing protocol and find the next hop because $F_{1}$ is the intended forwarder specified in the route. However, if $N_{1}$ receives the coded packet, since it was not specified in the route, it may not be able to find the correct next hop. Thus, $N_{1}$ as a non-intended forwarder drops coded packets.

FlexONC moves one step further for more diffusion gain than BEND, and allows non-intended forwarders to cooperate in receiving and forwarding of not only native packets but also coded packets. In fact, it provides the next-hop information of decoded packets to non-intended forwarders so that they are able to forward the packet to the correct next hop toward the destination. As we explained in the previous section, by doing so FlexONC is able to provide more diffusion gain and more coding opportunities, which leads to higher throughput in comparison to previous methods.

\section{Overview of FlexONC}
\subsection{Motivating Example}
\begin{figure}[ht]
\centering
\includegraphics[scale=0.48]{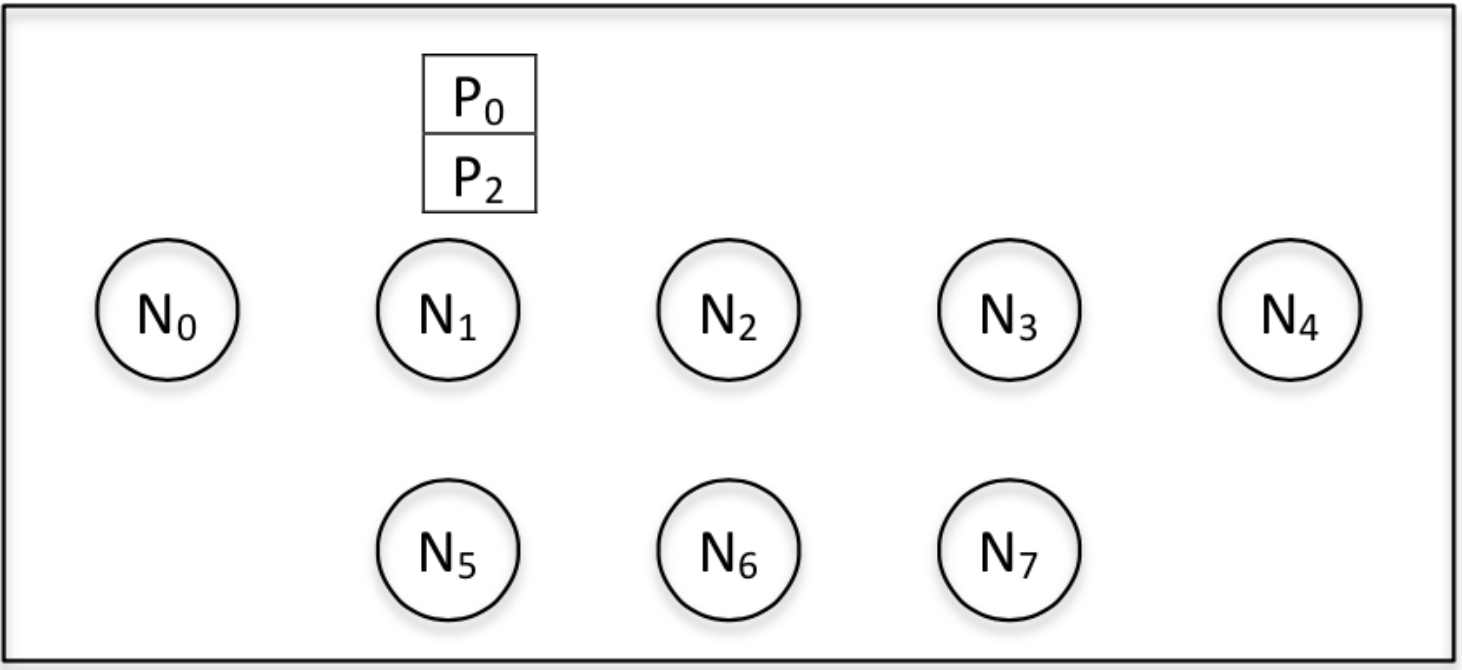}
\caption{Non-intended forwarders can help decoding.}
\label{8node}
\end{figure}

Fig. \ref{8node} presents an 8-node topology where there exist two flows from $N_{0}$ to $N_{4}$ and vice versa. We assume each node can hear only from nodes immediately next to it. As shown in this figure, $N_{1}$'s queue contains 2 native packets $P_{0}$ and $P_{2}$ with different next hops $N_{0}$ and $N_{2}$, respectively. Let us assume $P_{0}$'s next hop is $P_{2}$'s previous forwarder or one of its neighbors and vice versa. So, $N_{1}$ decides to mix these packets together, hoping that $N_{2}$ ($N_{0}$) has already received $P_{0}$ ($P_{2}$) and it is able to decode and retrieve its own packet $P_{2}$ ($P_{0}$). Therefore, $N_{1}$ sends a coded packet $P=P_{0} \oplus P_{2}$ to $N_{0}$ and $N_{2}$ (i.e., next-hop list in the packet header contains $N_{0}$ and $N_{2}$) while we assume $N_{6}$ overhears the packet. 

In the previous methods like COPE and BEND, $N_{6}$ discards the packet immediately either due to the fact that it is not the next hop (as in COPE) or because the packet is not a native packet (as in BEND). Here, we assume that $N_{2}$ does not receive the packet or has not received $P_{0}$ and cannot decode $P_{2}$, but $N_{6}$ receives it successfully, and also can decode the packet. In such a scenario, in previous methods, after a time-out, $N_{1}$, which has not heard any ACK from $N_{2}$, retransmits the packet. However, FlexONC avoids such unnecessary retransmissions and $N_{6}$ forwards the packet to its next hop on behalf of $N_{2}$.

In fact, FlexONC allows non-intended forwarders like $N_{6}$ to decode a received coded packet if they can, and forward it toward the final destination as long as they believe it can be beneficial. By doing so, since $N_{2}$ is not the only node in charge of forwarding packets, the traffic is spread in the network. That is if $N_{2}$ fails to receive or decode a packet, its role is immediately covered by $N_{6}$. This idea not only can accelerate packet delivery by removing some retransmissions but also can provide more coding opportunities. For example, let us further assume $N_{6}$ is going to forward $P_{2}$ on behalf of $N_{2}$. If $P_{2}$ is eligible to be mixed with some packets queued at $N_{6}$, by allowing $N_{6}$ to decode and forward it, we capture more coding opportunities in $N_{6}$.

\subsection{Objective and Challenges}
FlexONC should avoid unnecessary changes to the standard MAC protocols, and be as simple as possible to be feasible in real scenarios. Moreover, it should be compatible with different routing protocols despite few modifications. To realize such compatibility, while having more flexibility in forwarding and coding, FlexONC should address the following questions. 
\begin{itemize}
\item How to select the nodes that can help the intended forwarder to forward packets: In other words, how should we decide which nodes are eligible? For example, in Fig. \ref{8node}, when $N_{1}$ sends the packet, $N_{5}$, $N_{2}$ and $N_{6}$ may receive it, but are they good candidates to forward the packet? Which one has the first priority? 
\item Duplicate packets: Since more nodes cooperate to move packets toward the destination, their imperfect collaboration may cause a significant number of duplicate packets travelling in the network leading to unnecessary contention and collision. Some mechanisms are required to control duplicate packets in the network. 
\item Flexible forwarding but not too far from the specified route: Although in FlexONC, like BEND, packets may not follow the exact route specified by the routing protocol, we need to keep them around the determined route. To do so, BEND uses the \emph{second-next-hop} field in native packets. However, as we described earlier, it is not applicable to coded packets at non-intended forwarders. For example, in Fig. \ref{8node} when $N_{6}$ receives the coded packet, even if it can decode $P_{2}$, it does not know the address of next hop from $N_{2}$ toward the destination. Thus in FlexONC, to enable $N_{6}$ to forward this packet, a new approach is required so that non-intended forwarders can find the correct address of the next hop.
\end{itemize}
We address all these questions in the next section.

\section{Implementation Details}
As described earlier, the idea behind FlexONC is to have backup nodes which can decode and forward a packet in case that the intended forwarder fails to do so, either due to unsuccessful reception of the packet or lack of required packets in the buffer to decode the original packet. This section describes in detail the responsibility of the sender and the receiver of a coded packet to realize this idea, and answers the questions stated in the previous section.

\subsection{Participant Nodes in FlexONC}
\subsubsection{The Receiver}
If an intended forwarder (e.g., $N_{2}$ in Fig. \ref{8node}) receives a coded packet and can decode it, it simply sends an ACK to the source. In FlexONC, an ACK contains the address of its sender instead of the receiver, the same as in BEND. In this case, non-intended forwarders (e.g., $N_{6}$) hear the ACK realizing that the intended forwarder has decoded the packet successfully and does not need their help.

On the other hand, when a non-intended forwarder receives a coded packet, if it can decode the packet and it is eligible to forward the packet, it sets a timer for anticipated ACKs. If it hears an ACK for the decoded packet from one of the nodes with a higher priority before time-out, it drops the packet. Otherwise, it sends an ACK to the sender, mixes possibly the decoded packet with other packets in the queue, and forwards it.

\subsubsection{The Sender}
It is obvious that when a sender sends a combination of $n$ packets, it should wait to receive $n$ ACKs. Thus, its waiting time before time-out is more than when it transmits a native packet. In FlexONC, because more nodes can help in decoding and forwarding a packet, if the sender does not hear an ACK from the intended forwarder, there is still a chance that it receives the ACK from a non-intended forwarder. Therefore, the sender should wait a little longer before it retransmits the packet. In fact, in FlexONC the waiting time of the sender for coded packets is proportional to the number of neighbors of the sender instead of the number of mixed packets.

\subsection{How Non-intended Forwarders Know the Address of the Next Hop?}

In FlexONC, although packets may not follow the exact route specified by the routing protocol, they travel around it and do not stray too far away. Thus, when a non-intended forwarder forwards the packet on behalf of the intended forwarder, it should send it to the next hop toward the destination from the intended forwarder's point of view. For example in Fig. \ref{8node}, when $N_{1}$ sends the coded packet $P=P_{0} \oplus P_{2}$, $N_{0}$, $N_{5}$, $N_{2}$, and $N_{6}$ may receive the packet. If $N_{2}$, which is the intended next hop for $P_{2}$, fails to receive the packet successfully, and if one of the non-intended forwarders (e.g., $N_{5}$, $N_{0}$, $N_{6}$) wants to forward it, they need to know the address of the next hop from $N_{2}$ toward the destination (not from themselves), which is $N_{3}$ in this example.

Since the \emph{second-next-hop} field in BEND cannot solve this problem, instead of adding this field to the packet header, in FlexONC, the routing protocol is enhanced such that each node also maintains forwarding tables of all its neighbors. As such, when for example $N_{6}$ forwards $P_{2}$ on behalf of $N_{2}$, it knows the address of the next hop from $N_{2}$ toward the destination, and simply sends the packet to it.

\subsection{Who Is Eligible to Decode and Forward a Packet?}
When a sender transmits a coded packet, all of its neighbors may receive it. However, every node that receives the packet is not necessarily eligible to forward it. In addition, if all eligible nodes were to forward the same packet, that would be a huge waste of the network bandwidth as well as a source of collision. We need a method to prioritize the eligible nodes.

In FlexONC, when a node like $N_{6}$ in Fig. \ref{8node} receives a coded packet, it first looks for its address in the next-hop list. If it cannot find its address, clearly it is not the intended forwarder for any native packet mixed in the coded packet. Therefore, as a non-intended forwarder, $N_{6}$ searches for a native packet in the coded packet that 1) its intended forwarder (e.g., $N_{2}$ for $P_{2}$ in Fig. \ref{8node}) is $N_{6}$'s neighbor, 2) its next hop from the intended forwarder (e.g., $N_{3}$ for $P_{2}$ in Fig. \ref{8node}) is $N_{6}$'s neighbor, and 3) it is decodable by $N_{6}$. Based on these criteria, in Fig. \ref{8node}, although when $N_{1}$ sends coded packet $P$, $N_{0}$, $N_{5}$ and $N_{6}$ as well as $N_{2}$ may receive the packet, $N_{0}$ is not eligible to forward $P_{2}$ due to the first criterion. Furthermore, $N_{5}$ is not qualified for the second criterion and therefore $N_{6}$ is the only non-intended forwarder which can send $P_{2}$ on behalf of $N_{2}$ if it can decode it.

\begin{figure}[ht]
\centering
\includegraphics[scale=0.64]{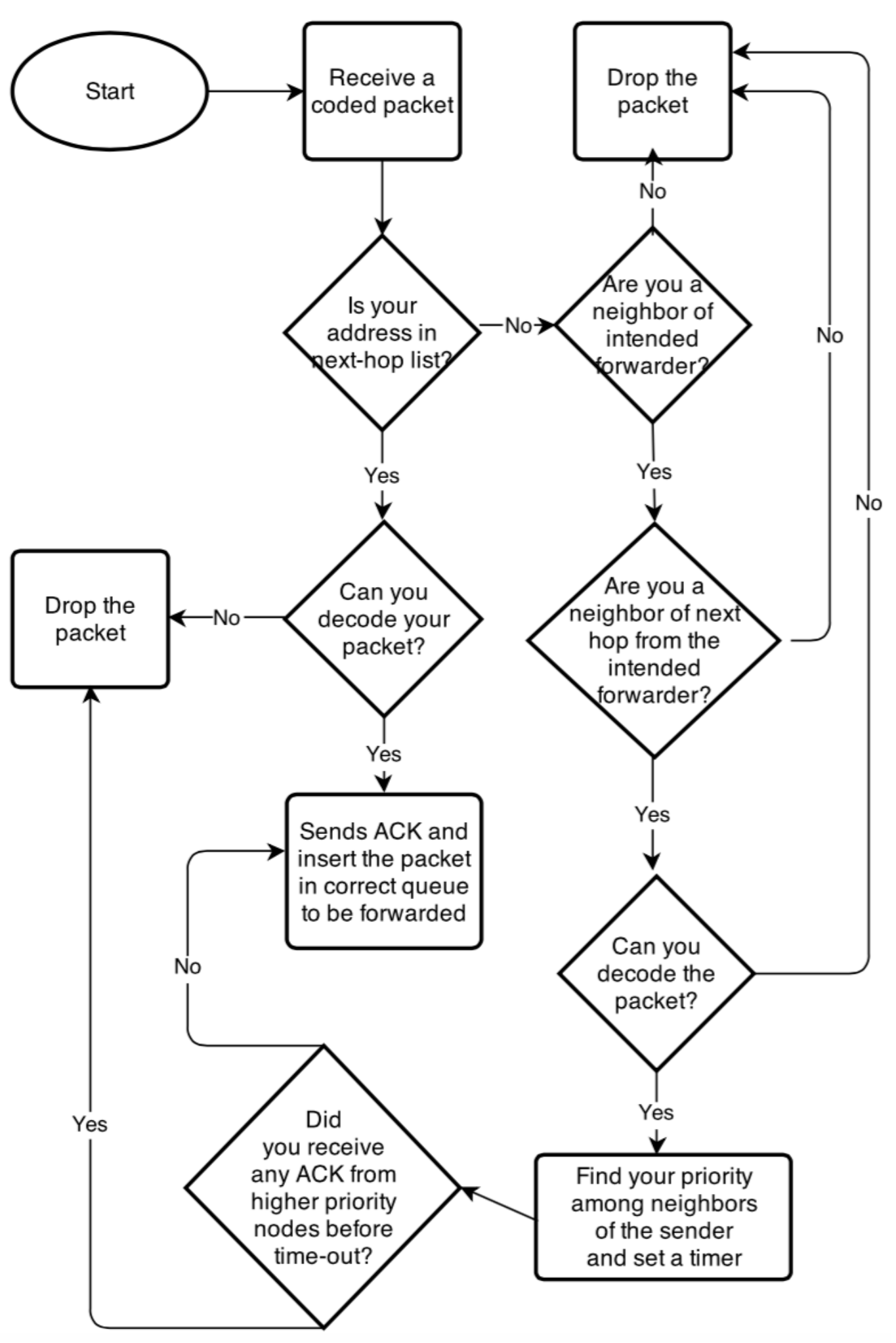}
\caption{Flowchart for receivers of coded packets in FlexONC.}
\label{flowChart}
\end{figure}

However, a non-intended forwarder should not forward a packet immediately after decoding it because the intended forwarder may forward the packet itself and does not need non-intended forwarders' help. In addition, if there are more than one eligible non-intended forwarders, an ordering among them is required. As such, in FlexONC when a non-intended forwarder receives a coded packet, it sorts the addresses of all neighbors of the sender (i.e., all non-intended forwarders and intended forwarders), gives the first priority to the intended forwarder of the decoded packet, and considers its index in the sorted list as its priority. Then it sets a timer and waits for an ACK from any node with a higher priority. If it does not hear any ACK after time-out, it is likely that none of the nodes with a higher priority has received and can forward the packet, so it is its turn to send the ACK back to the sender and forward the packet. Fig. \ref{flowChart} presents the flowchart for receivers of a coded packet in FlexONC.

\subsection{How to Limit the Number of Duplicate Packets?}
Although FlexONC aims to eliminate duplicate packets by prioritizing non-intended forwarders and making the sender wait for their ACK, duplicate packets may still exist in the network, due to several reasons such as lack of perfect synchronization. Therefore, FlexONC relies on more strategies to control the number of duplicate packets in the network. First, after receiving an ACK, if the node finds a packet in its buffer that the sender of the ACK is the next hop of the packet or one of its neighbors, the node drops the packet (i.e., the packet has already received by down stream nodes). Second, in FlexONC each node stores a limited number of received ACKs and if it receives a packet, it searches this ACK list. If it can find an ACK for the same packet sent by the next hop of the packet or one of its neighbors, it also drops the packet.

\section{Performance Evaluation}
We use Network Simulator ns-2 to compare the performance of FlexONC against BEND, a simulation version of COPE called COPE-Sim \cite{BEND} and IEEE 802.11\footnote{Note that FlexONC, BEND and COPE use the same data link layer signalling as IEEE 802.11, and thus, we consider 802.11 as one of the baselines in our performance evaluation.}. The rest of this section describes the experiment scenarios as well as the performance results in two different topologies.

\subsection{Network Description}
In our simulation, we use the physical layer model implemented by BEND, so that by choosing different BERs (bit error rates), we can study the performance under different link qualities and packet loss probabilities. Since the channel propagation used in ns-2 is a Two-Ray Ground Reflection model, the maximum transmission range is 250 m. Data rate is fixed to 1 Mbps. The sources, in our simulation scenarios, send CBR (i.e., Constant Bit Rate) data flows with datagram size of $1000$ bytes. Also, we use DSDV as the routing protocol and apply few minor changes so that each node can obtain forwarding tables from its neighbors.

To investigate the performance of FlexONC in comparison to BEND, COPE-Sim and 802.11, we test them in different scenarios and compare their throughput as well as the throughput gain of FlexONC over the baselines for different BERs in two topologies. First, we compare them using a simple 8-node topology shown in Fig. \ref{8node}, and then we use a $5 \times 5$ grid topology as a more general case.

\subsection{8-Node Topology}
In the 8-node topology presented in Fig. \ref{8node}, two flows in opposite directions transmit packets from $N_{0}$ to $N_{4}$ and vice versa. Since the distance between adjacent nodes in both $X$ and $Y$ axes is 150 m, each node can receive packets only from nodes immediately next to it horizontally, vertically, or diagonally (e.g., $N_{1}$ can hear from $N_{0}$, $N_{5}$, $N_{2}$, and $N_{6}$). The arrival interval of CBR flows in these scenarios is 0.07 s and its duration is 150 s. 

\begin{figure}[!t]
\centering
\includegraphics[scale=0.48]{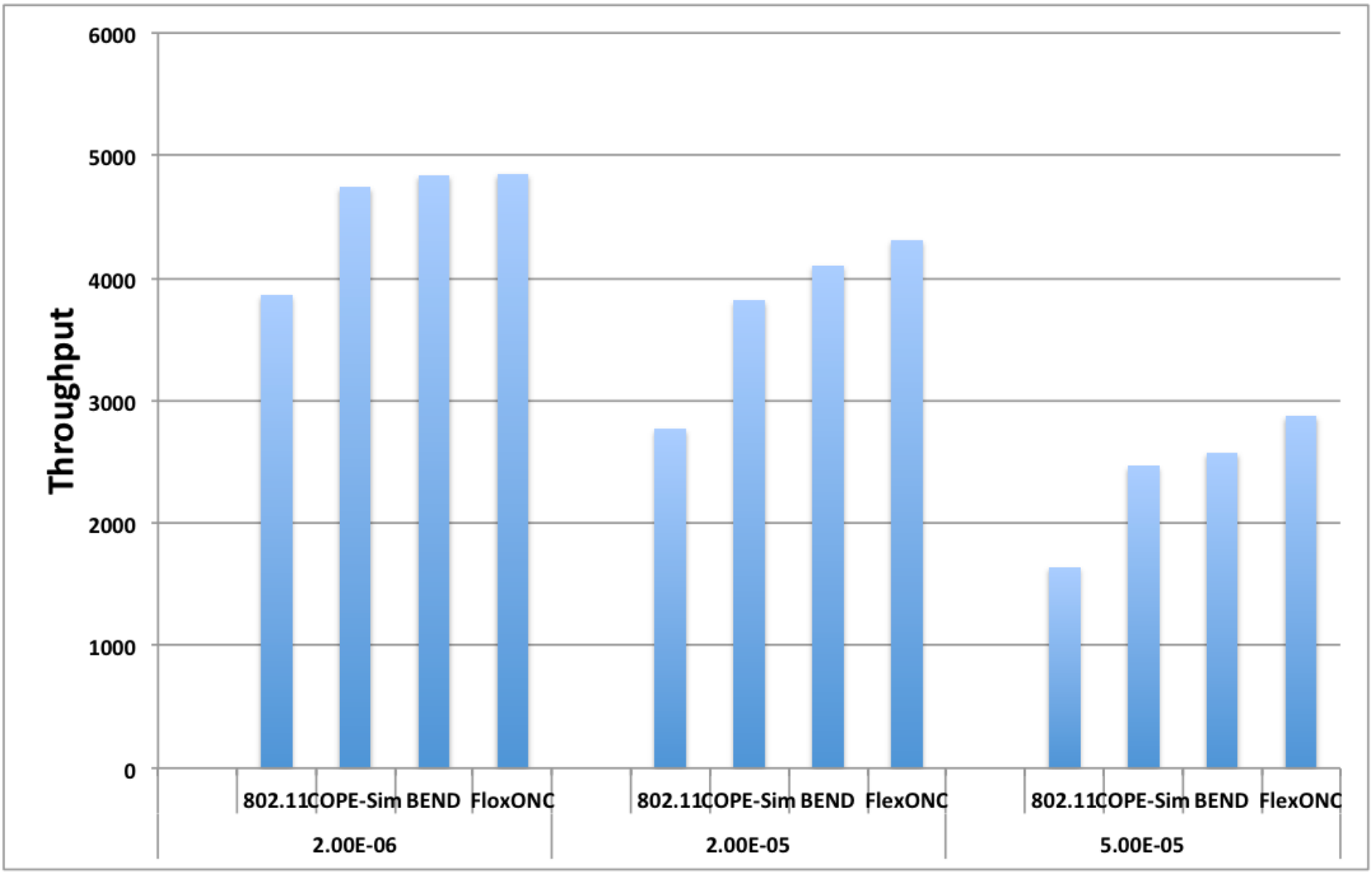}
\caption{Throughput of different methods in 8-node topology for different BERs.}
\label{throughput8}
\end{figure}

In this topology, for each intended forwarder except for the final destination, there exists at least one non-intended forwarder that can help the intended forwarder and take responsibility of forwarding packets when the intended forwarder fails to do so. Fig. \ref{throughput8} presents the throughput of BEND, COPE-Sim, and 802.11 as well as FlexONC for three lowest BERs in our experiments.

We can observe that when BER = $2 \times 10^{-6}$ (i.e., the network condition is almost perfect), most transmitted packets are received by the intended forwarders successfully. Therefore, there hardly exists an opportunity for non-intended forwarders to decode and forward a packet on behalf of the intended forwarder. It is obvious that in such a situation, FlexONC can not show its real power and its throughput is close to BEND. However, as the BER increases, more opportunities for non-intended forwarders are provided and FlexONC's gain over other methods increases significantly.

Table \ref{table:8node} presents the performance gain of FlexONC over BEND, COPE-Sim and 802.11 for 6 different BER levels, which corroborates our observation. In particular, by increasing the BER, FlexONC becomes more powerful in comparison to the baselines, and its throughput gain increases.
   
\begin{table}[!t]
\renewcommand{\arraystretch}{1.3}
\caption{FlexONC's gain over other methods in 8-node topology.}
\label{table:8node}
\centering
\begin{tabular}{|c||c||c||c|}
\hline
\textbf{BER} & \textbf{BEND} & \textbf{COPE-Sim} & \textbf{802.11}\\ \hline
\textbf{$2 \times 10^{-6}$} &	0.2$\%$  &	2$\%$ & 26$\%$\\ \hline
\textbf{$2 \times 10^{-5}$} &	5$\%$ &	13$\%$ & 56$\%$\\ \hline
\textbf{$5 \times 10^{-5}$} &	12$\%$ &	 17$\%$ & 75$\%$\\ \hline
\textbf{$8 \times 10^{-5}$} &	13$\%$ &	 21$\%$ & 89$\%$\\ \hline
\textbf{$1 \times 10^{-4}$} &	26$\%$ &	 34$\%$ & 158$\%$\\ \hline
\textbf{$2 \times 10^{-4}$} &	60$\%$ &	 86$\%$ & 265$\%$\\ \hline
\end{tabular}
\end{table}

\subsection{Mesh Topology}
\begin{figure}[!t]
\centering
\includegraphics[scale=0.46]{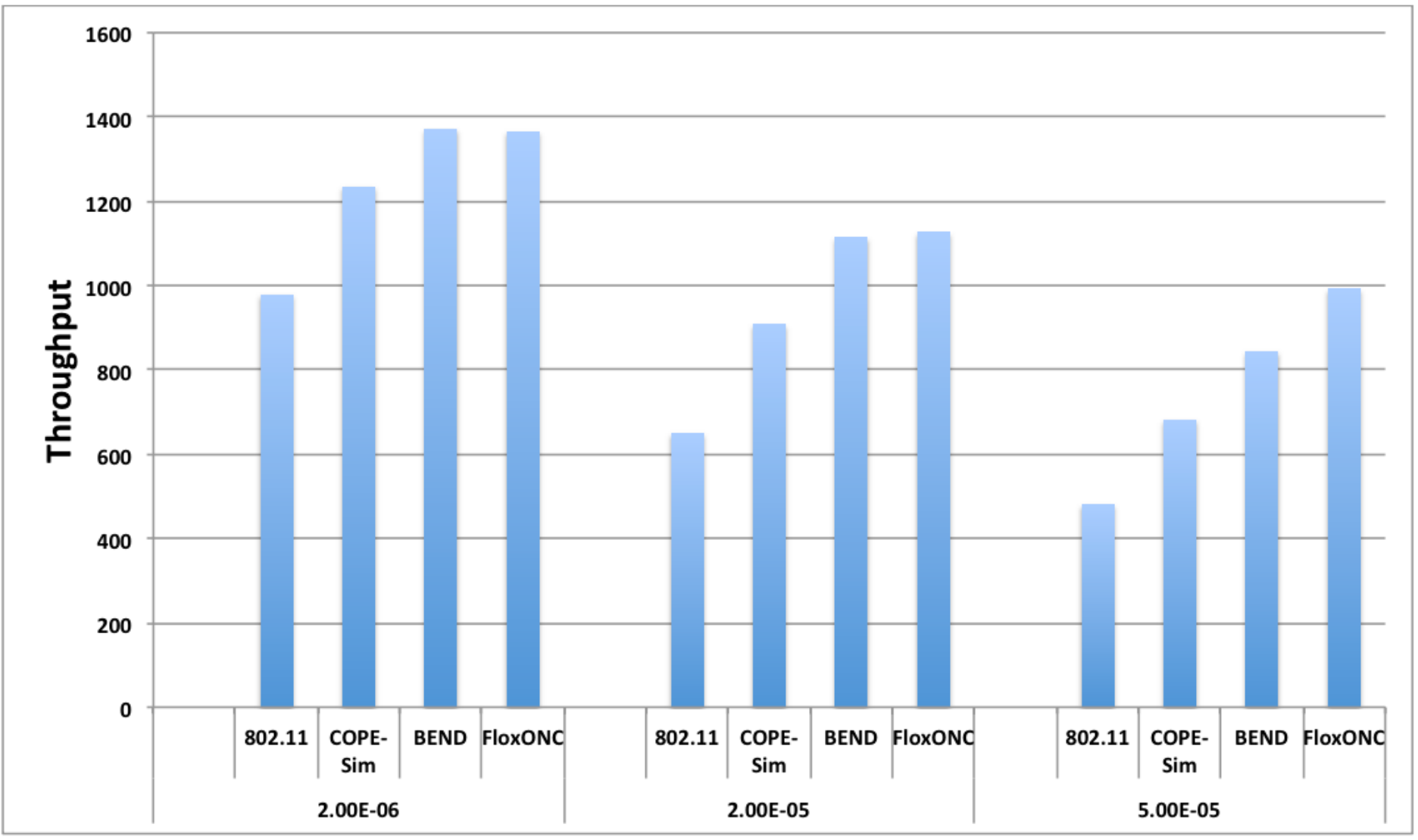}
\caption{Throughput of different methods in mesh topology for different BERs.}
\label{throughputMesh}
\end{figure}

To investigate the performance of FlexONC in a general topology, we test it in a $5 \times 5$ grid where again the distance between two adjacent nodes in both axes is 150 m. 8 different flows with arrival interval of 0.1 s and duration of 100 s transmit packets between Row 1 and Row 5, and also Column 1 and Column 5 of the grid. The performance result depicted in Fig. \ref{throughputMesh} and Table \ref{table:mesh} again shows that FlexONC almost always outperforms BEND and other methods especially at non-trivial BER level.

\begin{table}[!t]
\renewcommand{\arraystretch}{1.3}
\caption{FlexONC's gain over other methods in mesh topology.}
\label{table:mesh}
\centering
\begin{tabular}{|c||c||c||c|}
\hline
\textbf{BER} & \textbf{BEND} & \textbf{COPE-Sim} & \textbf{802.11}\\ \hline
\textbf{$2 \times 10^{-6}$} &	$-0.3\%$  &	10$\%$ & 39$\%$\\ \hline
\textbf{$2 \times 10^{-5}$} &	1$\%$ &	24$\%$ & 73$\%$\\ \hline
\textbf{$5 \times 10^{-5}$} &	17$\%$ &	 45$\%$ & 106$\%$\\ \hline
\textbf{$8 \times 10^{-5}$} &	29$\%$ &	 47$\%$ & 161$\%$\\ \hline
\textbf{$1 \times 10^{-4}$} &	40$\%$ &	 53$\%$ & 174$\%$\\ \hline
\textbf{$2 \times 10^{-4}$} &	52$\%$ &	 69$\%$ & 156$\%$\\ \hline
\end{tabular}
\end{table}
\section{Discussion}
In our experiments, we selected DSDV as the routing protocol since it is a well-know protocol. Moreover, it is a distance-vector approach that makes fewer assumptions about the routing information in comparison to source routing protocols. Therefore, if FlexONC works well with DSDV, it will work with source routing protocols as well. As a matter of fact, choosing DSDV as the routing module does not lose generality in our idea, and we believe it would not make a big difference in FlexONC's performance gain if we chose any other routing protocol as long as the routing protocol can be modified in a way that each node contains forwarding information for its neighbors.

On one hand, FlexONC decreases the delay in forwarding packets and increases the throughput by avoiding packet retransmission when an intended forwarder fails to decode the coded packet, and a non-intended forwarder alternatively pass the packet toward the destination. On the other hand, when more nodes have the responsibility of passing the packet further to the destination, the sender should wait longer for an ACK before it retransmits the packet, and this longer waiting time means longer delay which may lead to lower throughput. 

Therefore, we face a trade off here. While the maximum waiting time of the sender is proportional to the number of its neighbors, the gain over FlexONC is also related to the number of neighbors of the sender (i.e., to be said more precisely, non-intended forwarders), as well as the probability of intended forwarder's failure in receiving or decoding a coded packet, which is in turn affected by the packet loss probability and BER in the network. The performance result showed that even for a very low BER that the intended forwarder itself can decode and forward the majority of received coded packets and FlexONC's idea does not have much chance to be applied, its performance is comparable to BEND.

\section{Conclusion and Future Work}
This paper presented FlexONC (Flexible and Opportunistic Network Coding), an enhancement over BEND, which provides more flexibility and coding opportunities in the network. By utilizing the broadcasting nature of wireless networks, FlexONC is able to spread different flows better than BEND and enable a higher level of cooperation between intended and non-intended forwarders at the link layer in a multi-hop wireless network.

The performance results show that at higher bit error rates, when an intended forwarder may fail to receive or decode a coded packet and needs its neighbor's help, FlexONC significantly outperforms previous methods like BEND, COPE and 802.11. Even under an ideal network condition, when intended forwarders usually do not need any help and can decode and forward received coded packets, FlexONC's performance is as good as BEND. 

In general, network coding significantly supports UDP flows, but for TCP flows, it may achieve much lower than expected gain because of the congestion control mechanism in TCP windows. However, in recent years little research has been conducted to control sent and received packets and ACKs to the transport layer, so that network coding can be applied without much effect on TCP windows \cite{TCP/NC, TCPNC2010}. Hence, a future extension of FlexONC could be its exploration and modification under TCP flows.

In addition, in recent years a few publications have been presented that apply both inter- and intra-session network coding, but in some limited scenarios \cite{I2NC, CORE1, CORE2}. We believe that this combination, if realized carefully, could introduce further improvement in the performance, and represents another way to extend FlexONC.

\bibliographystyle{IEEEtran} 
\bibliography{citation} 

\end{document}